\documentclass{PoS}
\usepackage{ulem}
\title{Probing the nature of Dark Matter with the SKA}
\ShortTitle{Unveiling Dark Matter with SKA}
\author{
\speaker{Sergio Colafrancesco}$^1$, 
Marco Regis$^{2}$,
Paolo Marchegiani$^{1}$,
Geoff Beck$^{1}$,
Rainer Beck$^{3}$,
Hannes Zechlin$^{2}$,
Andrei Lobanov$^{3}$,
Dieter Horns$^{4}$
\\ 
$^1$School of Physics, University of the Witwatersrand, Johannesburg, South Africa\\
$^2$Dipartimento di Fisica, Universit\`{a} degli Studi di Torino and INFN-Sezione di Torino, \\ via P. Giuria, 1, 10125 Torino, Italy\\
$^3$Max-Planck-Institut f\"ur Radioastronomie, Auf dem H\"ugel 69, 53121 Bonn, Germany\\
$^4$Institut f\"ur Experimentalphysik, Universit\"at Hamburg, Luruper Chaussee 149, 22761 Hamburg, Germany\\
\\
E-mail: \email{sergio.colafrancesco@wits.ac.za}
}

\abstract{
Dark Matter (DM) is a fundamental ingredient of our Universe and of structure formation, and yet its nature is elusive to astrophysical probes.  Information on the nature and physical properties of the WIMP (neutralino) DM (the leading candidate for a cosmologically relevant DM) can be obtained by studying the astrophysical signals of their annihilation/decay. Among the various e.m. signals, secondary electrons produced by neutralino annihilation generate synchrotron emission in the magnetized atmosphere of galaxy clusters and galaxies which could be observed as a diffuse radio emission (halo or haze) centered on the DM halo.  
A deep search for DM radio emission with SKA in local dwarf galaxies, galaxy regions with low star formation and galaxy clusters (with offset DM-baryonic distribution, like e.g. the Bullet cluster) can be very effective in constraining the neutralino mass, composition and annihilation cross-section.  For the case of a dwarf galaxy, like e.g. Draco, the constraints on the DM annihilation cross-section obtainable with SKA1-MID will be at least a factor $\sim 10^3$ more stringent than the limits obtained by Fermi-LAT in the $\gamma$-rays. These limits scale with the value of the B field, and the SKA will have the capability to determine simultaneously both the magnetic field in the DM-dominated structures and the DM particle properties.  The optimal frequency band for detecting the DM-induced radio emission is around $\sim 1$ GHz, with the SKA1-MID Band 1 and 4 important to probe the synchrotron spectral curvature at low-$\nu$ (sensitive to DM composition) and at high-$\nu$ (sensitive to DM mass). 
}

\FullConference{Advancing Astrophysics with the Square Kilometre Array,\\
		June 8-13, 2014\\
		Giardini Naxos, Sicily, Italy }

\newcommand{\skipthis}[1]{}

\def\simlt{\ \raise -2.truept\hbox{\rlap{\hbox{$\sim$}}\raise5.truept   %
\hbox{$<$}\ }}
\def\simgt{\ \raise -2.truept\hbox{\rlap{\hbox{$\sim$}}\raise5.truept   %
\hbox{$>$}\ }}                                                          %

\begin{document}

\section{Unveiling the nature of Dark Matter}

The Square Kilometre Array (SKA) is the most ambitious radio telescope ever planned, and it is a unique multi-disciplinary experiment. Even though the SKA, in its original conception, has been dedicated to constrain the fundamental physics aspects on dark energy, gravitation and magnetism, much more scientific investigation could be done with its configuration: the exploration of the nature of Dark Matter is one of the most important additional scientific themes.

\begin{figure}[ht!]
\centering
\hbox{
\includegraphics[scale=0.55]{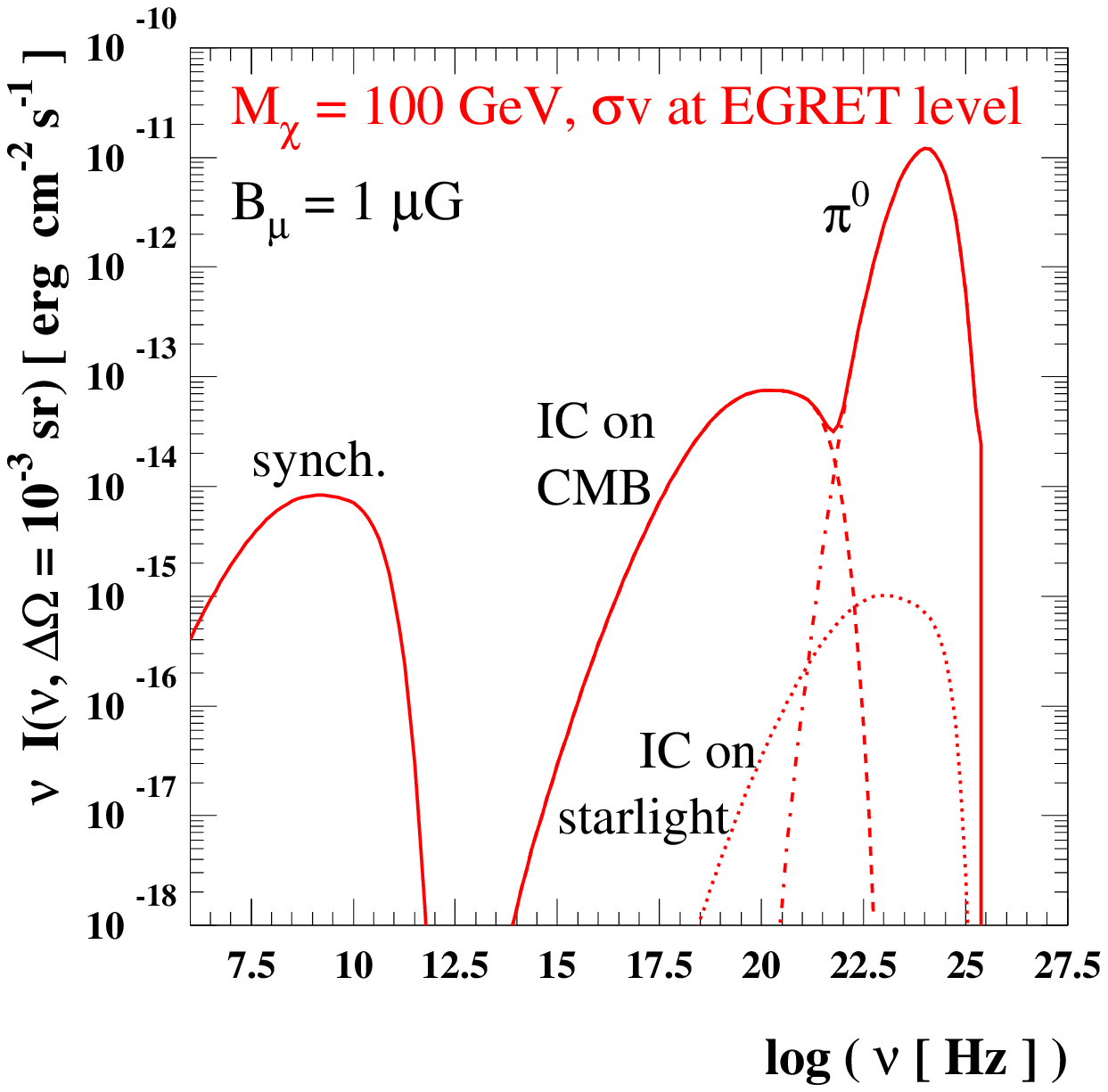}
\includegraphics[scale=0.55]{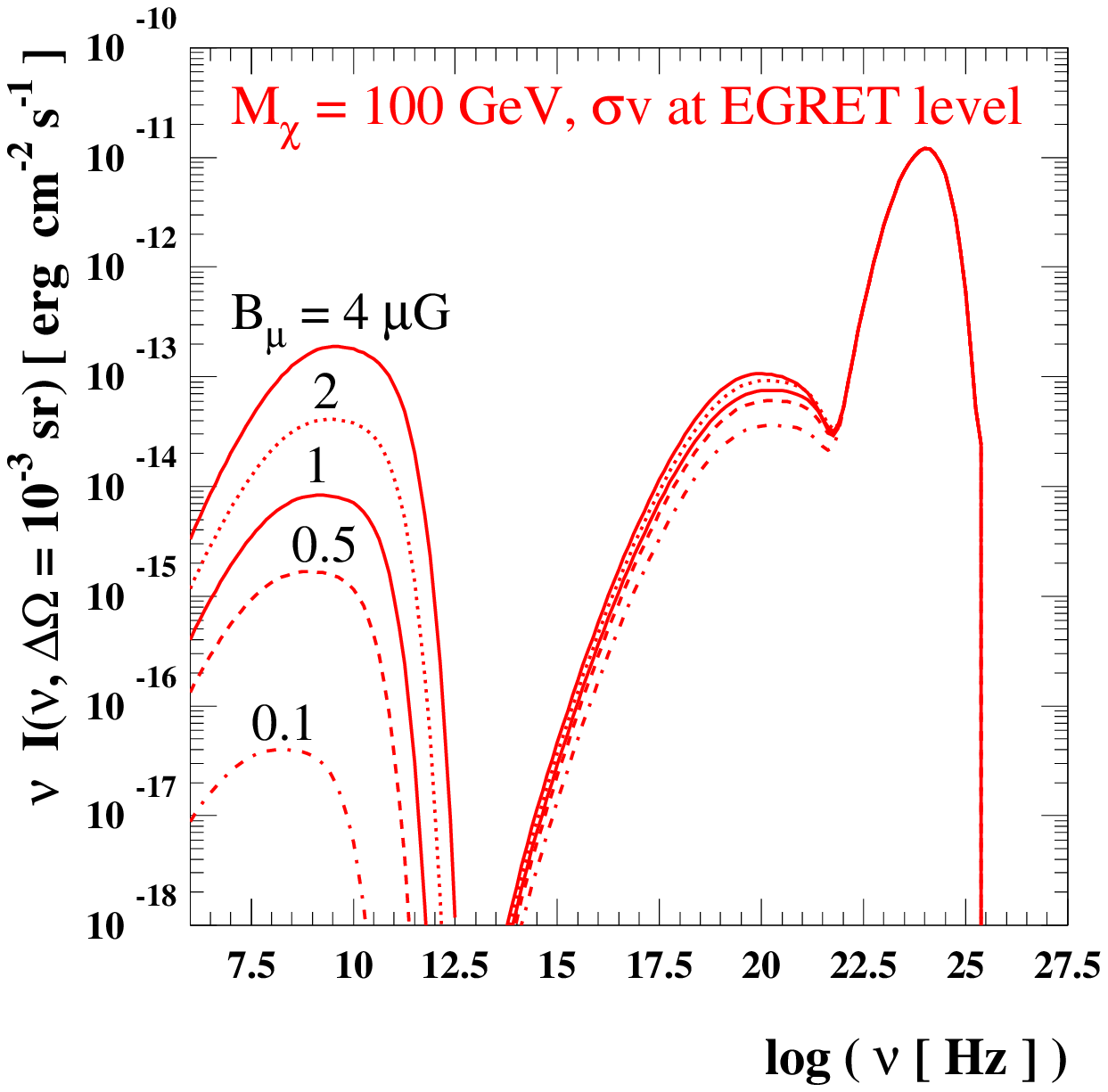}
}
\caption{Left. The Draco dSph multi-wavelength spectrum for a 100 GeV WIMP annihilating into
$b\bar{b}$. Right. The effect of varying the magnetic field strength on the Draco multi-wavelength spectrum for a 100 GeV WIMP annihilating into $b\bar{b}$. The WIMP pair annihilation rate has been tuned as to give a $\gamma$-ray signal at the level of the EGRET measured flux upper limit (from Colafrancesco et al. 2007).}
\label{fig:multiflux}
\end{figure}

Among the viable competitors for having a cosmologically relevant
DM species, the leading candidate is the lightest particle of the
minimal supersymmetric extension of the Standard Model (MSSM, Jungman et al. 1996), plausibly the neutralino $\chi$, with a mass $M_{\chi}$ in the range between a few GeV to several TeV.
Information on the nature and physical properties of the neutralino DM can be obtained by studying the astrophysical
signals of their interaction/annihilation in the halos of cosmic structures. These signals (see Colafrancesco et al. 2006, 2007 for details)
involve, in the case of a $\chi$ DM, emission along a wide range of frequencies, from radio to $\gamma$-rays (see Fig. \ref{fig:multiflux} for a DM spectral energy distribution (SED) in a typical dwarf galaxy).
Neutral pions produced in $\chi \chi$ annihilation decay promptly in $\pi^0 \to \gamma \gamma$ and generate most of the continuum photon spectrum at energies $E \gtrsim 1$ GeV.
Secondary electrons are produced through various prompt generation mechanisms and by the decay of charged pions, $\pi^{\pm}\to
\mu^{\pm} + \nu_{\mu}(\bar{\nu}_{\mu})$, with $\mu^{\pm}\to e^{\pm} + \bar{\nu}_{\mu}(\nu_{\mu}) + \nu_e (\bar{\nu}_e)$.
The different composition of the $\chi\chi$ annihilation final state will in general affect the form of the electron spectrum.
The time evolution of the secondary electron spectrum is described by the transport equation:
\begin{equation}
\frac{\partial n_e}{\partial t} = \nabla \left[ D \nabla
n_e\right] + \frac{\partial}{\partial E} \left[ b_e(E) n_e
\right]+ Q_e(E,r)\;,
 \label{diffeq}
\end{equation}
where $Q_e(E,r)$ is the $e^{\pm}$ source spectrum, $n_e(E,r)$ is the $e^{\pm}$ equilibrium spectrum (at each fixed time)
and $b_e\equiv -dE/dt$  is the $e^{\pm}$ energy loss per unit time, $b_e =b_{ICS} + b_{synch} + b_{brem} + b_{Coul}$ (see Colafrancesco et al. 2006 for details).
The diffusion coefficient $D$ in eq.(\ref{diffeq}) sets the amount of spatial diffusion for the secondary electrons: it turns out
that diffusion can be neglected in galaxy clusters while it is relevant on galactic and sub-galactic scales (see discussion in Colafrancesco et al. 2006, 2007).
Under the assumption that the population of high-energy $e^{\pm}$ can be described by a quasi-stationary ($\partial n_e / \partial t \approx 0$) transport equation, the secondary electron spectrum $n_e (E,r)$ reaches its equilibrium configuration mainly due to
synchrotron and ICS losses at energies $E \gtrsim 150$ MeV and due to Coulomb losses at lower energies.
Secondary electrons eventually produce radiation by synchrotron emission in the magnetized atmosphere of cosmic structures, bremsstrahlung with ambient protons and ions, and ICS of CMB (and other background) photons (and hence an SZ effect, Colafrancesco 2004). These secondary particles also heat the ambient gas by Coulomb collisions. 

A large amount of efforts have been put in the search for DM indirect signals at $\gamma$-ray energies looking predominantly for two key spectral features: the $\pi^0 \to \gamma \gamma$ decay spectral bump , and the direct $\chi \chi \to \gamma \gamma$ annihilation line emission, with results that are not conclusive yet (e.g., Daylan et al. 2014, Weniger 2012, Doro et al. 2014). 
The non-detection of signals related to DM annihilation/decay from various astrophysical targets (including observations of dwarf spheroidal galaxies, the Galactic Center, galaxy clusters, the diffuse gamma-ray background emission) is interpreted in terms of constraints on the (self-)annihilation cross-section (or decay time) of the DM particle candidate. Assuming, for instance, a canonical WIMP of $M_{\chi}=100$\,GeV annihilating to $b$-quarks, stacked observations of dwarf spheroidal galaxies with Fermi-LAT put a constraint of $\langle \sigma V \rangle < 2\times 10^{-25}\,\mathrm{cm}^3\,\mathrm{s}^{-1}$ (95\% c.l.) on the thermally-averaged self-annihilation cross-section of DM particles (Ackerman et al. 2014). Hopes of discovering annihilating WIMPs in $\gamma$-rays are relegated to Fermi-LAT successors and the forthcoming CTA experiment (Doro et al. 2013).\\
There are, however, also good hopes to obtain relevant information on the nature of DM from radio observations of DM halos on large scales, i.e. from dwarf galaxies to clusters of galaxies.

\section{Radio emission signals from Dark Matter annihilation}

Observations of radio halos produced by DM annihilations are, in principle, very effective in constraining the
neutralino mass and composition (see, e.g., Colafrancesco and Mele 2001, Colafrancesco et al. 2006, 2007),  under the hypothesis that DM annihilation provides an observable contribution to the radio-halo flux.


\begin{figure}[t!]
\centering
\includegraphics[scale=0.38]{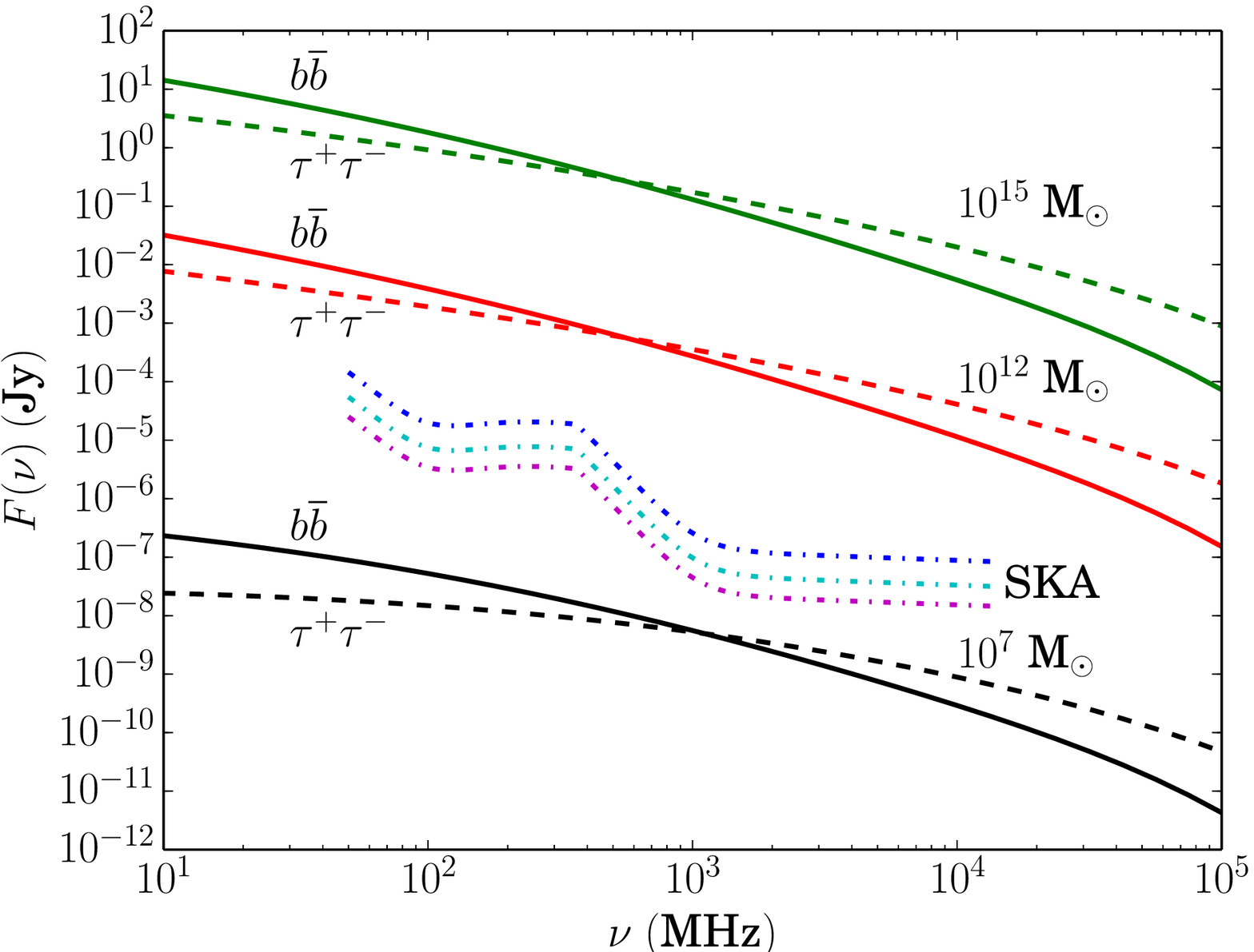}
\includegraphics[scale=0.38]{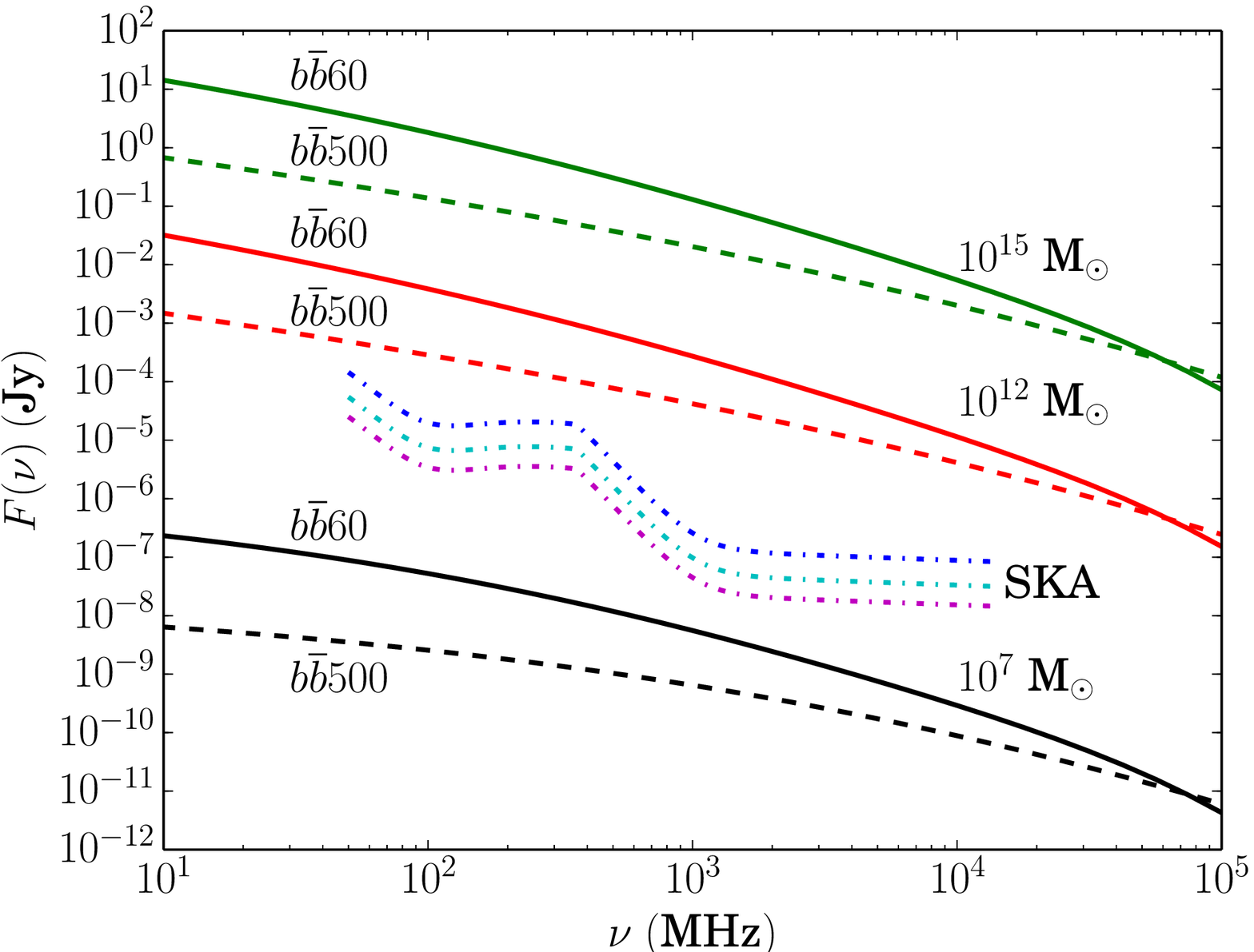}
\caption{Flux densities for dwarf galaxies ($M = 10^{7}$ M$_{\odot}$), galaxies ($M = 10^{12}$ M$_{\odot}$), and galaxy clusters ($M = 10^{15}$ M$_{\odot}$). The halo profile is NFW, $\langle B \rangle = 5$ $\mu$G, and the annihilation cross section is fixed to the value $\langle \sigma V \rangle \approx 3 \times 10^{-27}$ cm $^3$ s$^{-1}$. Black lines are for dwarf galaxies, red lines for galaxies and green for clusters. Left: WIMP mass is 60 GeV, solid lines are for composition $b\overline{b}$ and dashed lines for $\tau^+\tau^-$. Right: Composition is $b\overline{b}$, solid lines are for WIMP mass $60$ GeV and dashed lines for $500$ GeV. Dash-dotted lines are SKA sensitivity limits for integration times of 30, 240 and 1000 hours (Dewdney et al. 2012). The flux is calculated within the virial radius (from  Colafrancesco et al. 2014).}
\label{fig:spec_sig}
\end{figure}

The wide range of frequencies probed by the SKA and the variety of achievable observational targets (and in turn of magnetic fields) will allow testing the non-thermal electron spectrum from about 1 GeV to few hundreds of GeV, that is the most relevant range in the WIMP search.
Figure~\ref{fig:spec_sig} displays the predicted spectral differences between various annihilation channels and WIMP masses: note that these differences manifest mainly in low-$\nu$ slope variation for differing annihilation channels and in the high-$\nu$ spectral flattening/steepening for larger/smaller masses. The SKA sensitivity curves  for  SKA1-LOW and SKA1-MID are taken from Dewdney et al. (2012).\\
The surface brigthness produced by DM-induced synchrotron emission is heavily affected by diffusion  in small scale structures, e.g., dwarf and standard galaxies, while it is less important in large structures, e.g., galaxy clusters (see Colafrancesco et al. 2006-2007 for a detailed discussion).


Polarization from DM-induced radio emission is expected at very low fractional levels due to the fact that DM spatial and velocity distribution is nearly homogeneous and that DM annihilation is mediated by secondary particle production.
Therefore, a low polarization level of detectable radio signals in the directions of DM halos would be consistent with the DM origin of such radio emission.
Residual high-polarization signals could be hence attributed to astrophysical sources in the direction or within the DM halos, and one could use these signals to infer properties of the magnetic field in these structures (see R. Beck et al. 2015, and F. Govoni et al. 2015).

\subsection{Cosmological evolution of Dark Matter radio emission}

\begin{figure}[ht!]
\centering
\hbox{
\includegraphics[scale=0.27]{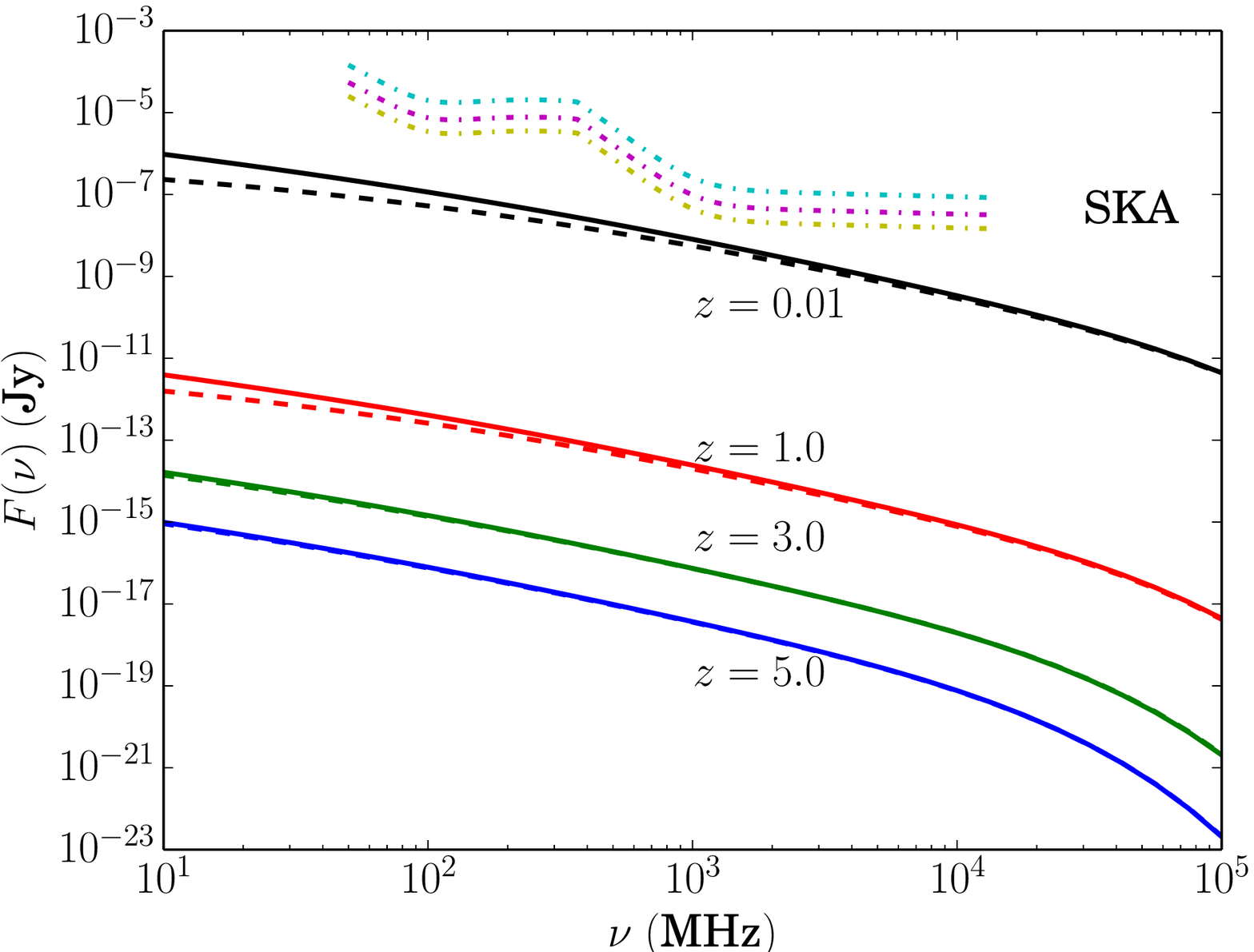}
\includegraphics[scale=0.27]{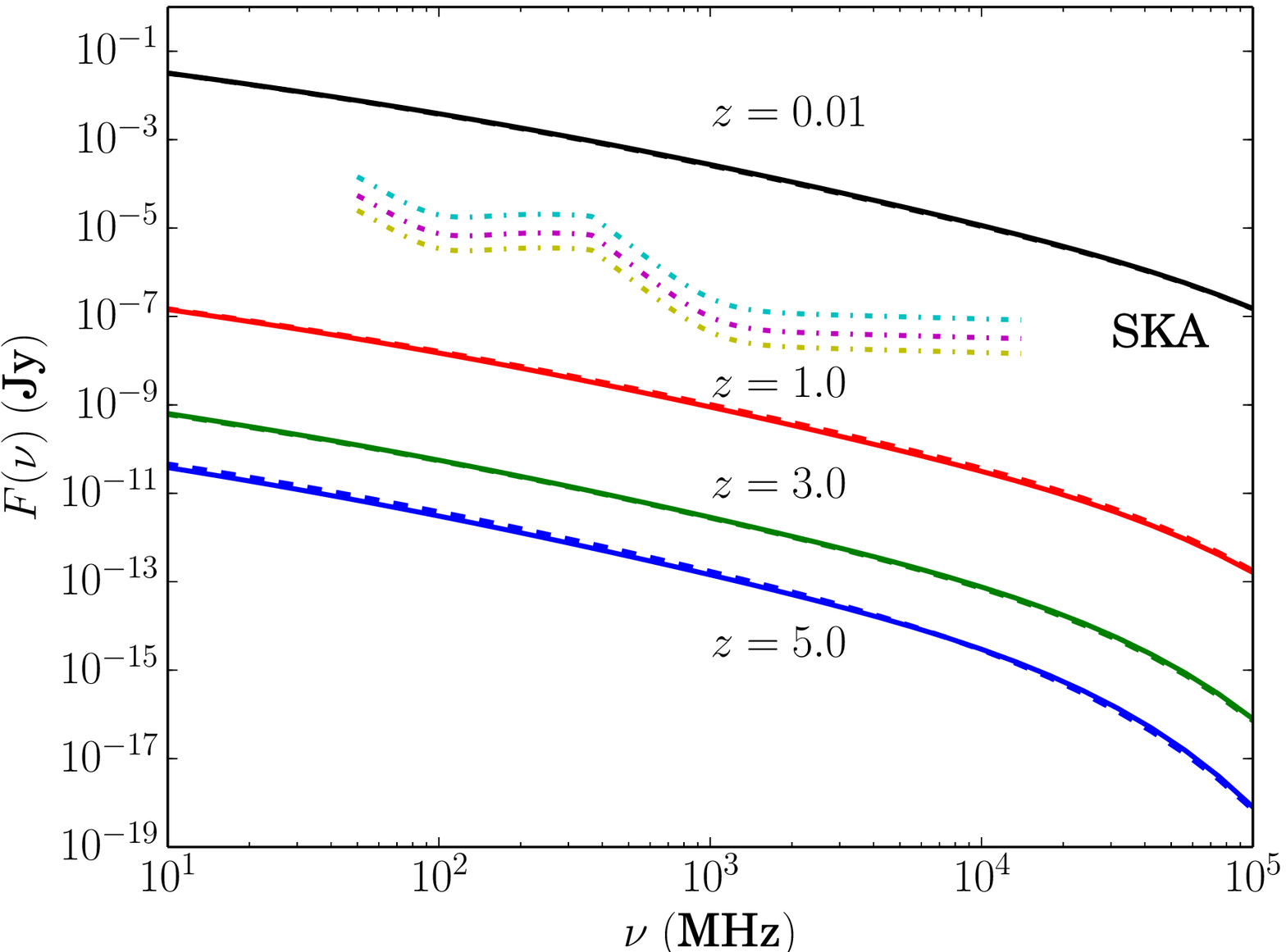}
\includegraphics[scale=0.27]{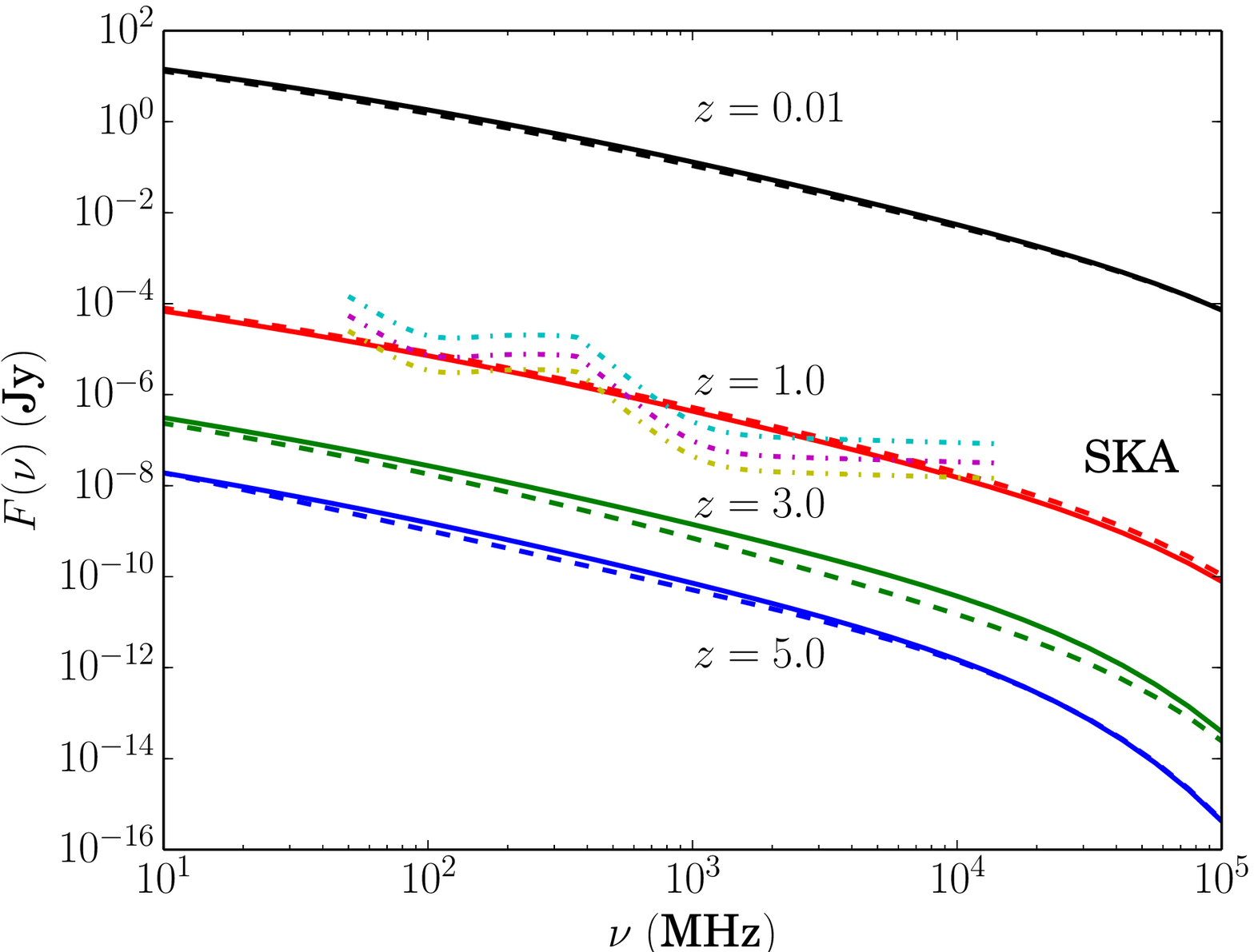}
}
\caption{Flux densities for various DM halos within the virial radius. Left: dwarf spheroidal galaxies ($M = 10^{7}$ M$_{\odot}$);  mid panel: galactic halos ($M = 10^{12}$ M$_{\odot}$); right panel: galaxy clusters ($M = 10^{15}$ M$_{\odot}$). We assume a NFW  halo profile, $\langle \sigma V \rangle = 3 \times 10^{-27}$ cm $^3$ s$^{-1}$ and $\langle B \rangle = 5$ $\mu$G. WIMP mass is $60$ GeV and the composition is $b\overline{b}$. Solid lines are without diffusion, and dotted are with diffusion.  The SKA sensitivity limits (dash-dotted lines, Dewdney et al. 2012) are shown for integration times of 30, 240 and 1000 hours, respectively (from Colafrancesco et al. 2014).
}
\label{fig:evol_bb60}
\end{figure}

Figure~\ref{fig:evol_bb60} displays the evolution of radio emission from DM halos of mass $10^{7}$ M$_{\odot}$, $10^{12}$ M$_{\odot}$, and $10^{15}$ M$_{\odot}$ for a constant magnetic field of $5$ $\mu$G, in accordance with arguments made in Colafrancesco et al. (2014). The annihilation channel is $b\overline{b}$, the mass of the neutralino is 60 GeV and a DM annihilation cross-section $\langle \sigma V \rangle = 3 \times 10^{-27}$ cm $^3$ s$^{-1}$ was adopted. Emission from dwarf galaxy halos ($M \approx 10^{7}$ M$_{\odot}$) is just below the SKA detection threshold for this value of  $\langle \sigma V \rangle$ but would be visible at redshift $z \leq 0.01$ for the assumed DM annihilation cross-section. This justify the search of DM-induced radio signals mainly in dwarf galaxies of the local environment, at distances $\simlt 3$ Mpc ($z \simlt 0.0007$).
Emission from galactic DM halos ($M \approx 10^{12}$ M$_{\odot}$) are detectable by SKA out to $z \approx 0.8$ even with the reference value of $\langle \sigma V \rangle$, and can provide a non-detection upper-bound on $\langle \sigma V \rangle$ an order of magnitude below the assumed value even at such high redshifts. 
Emission form galaxy cluster halos can  provide similar constraints but out to higher redshifts $z \simlt 3$. These objects thus offer the option of deep-field observations that can scan a larger fraction of the DM parameter space than the best current data.\\
Figure~\ref{fig:evol_bb60} also shows that the effect of diffusion are far less significant when observing higher-$z$ objects, again simplifying the modelling and analysis of the SKA observations.

\subsection{Optimal DM laboratories}

In order to identify the optimal DM laboratories for radio observations we scan a parameter space extending from dwarf galaxies to galaxy clusters over a wide redshifts range  $z \approx 0-5$. The choice to examine the halos of both large and small structures is crucial, as dwarf spheroidal galaxies are well known to be highly DM dominated but produce faint emissions, while larger structures, but not immaculate test-beds for DM emissions, provide substantially stronger fluxes. This indicates that a survey of DM halos with different mass is essential to identify the best detection prospects for future radio telescopes like the SKA.

\begin{figure}[ht!]
\centering
\includegraphics[scale=0.4]{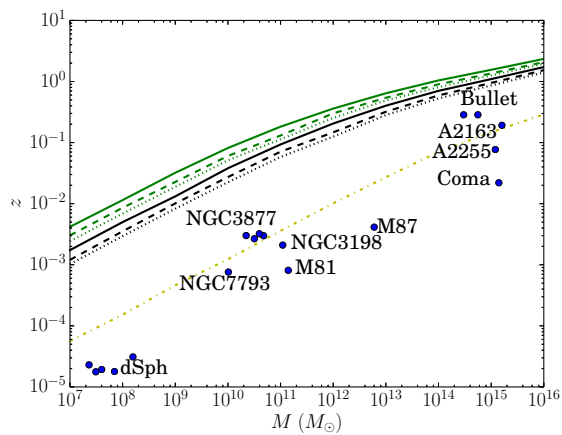}
\caption{Exclusion plot in redshift versus halo mass based on projected SKA (at 1 GHz in Band 1) sensitivity data for the reference value of $\langle \sigma V \rangle = 3 \times 10^{-27}$ cm $^3$ s$^{-1}$ with 30 hour integration time (black lines) and 1000 hour integration time (green lines). $\langle B \rangle = 5$ $\mu$G was adopted. Solid lines are the $1\sigma$ sensitivity exclusion, dashed lines that of $2\sigma$ and dotted lines correspond to $3\sigma$. The yellow dash-dotted line corresponds to 30 hours of integration and 1$\sigma$ confidence with $\langle \sigma V \rangle = 3 \times 10^{-30}$ cm$^3$ s$^{-1}$. An annihilation channel $b\overline{b}$ is assumed with a neutralino mass of 60 GeV. Representative objects with known DM mass are shown for illustrative purposes of DM radio signal detection. The dSph group contains the galaxies: Draco, Sculptor, Fornax, Carina and Sextans. Unlabelled Galaxies are: NGC3917, NGC3949 and NGC4010. 
For very local objects the redshift is estimated from the average distance data.
From Colafrancesco et al. 2014.}
\label{fig:loop}
\end{figure}

Figure~\ref{fig:loop} shows the redshift-mass exclusion plot obtained by using the SKA sensitivity bound for SKA1 LOW and SKA1 MID (at 1 GHz in Band 1). For each DM halo we obtain the DM halo mass and redshift combination that produce the minimal SKA-detectable fluxes. DM-dominated objects lying above the black and green curves cannot be detected with the SKA1 at the given confidence threshold for $\langle \sigma V \rangle = 3 \times 10^{-27}$ cm$^3$ s$^{-1}$. Objects below a curve are visible to SKA1, and the further below the curve they lie the greater the region of the cross-section parameter space we can explore through the observation of the object. For reference, the yellow dash-dotted line displays the curve given for $\langle \sigma V \rangle = 3 \times 10^{-30}$ cm$^3$ s$^{-1}$ and $1\sigma$ confidence level. A few representative know objects (irrespective of their location in the sky) with good estimates of the DM mass are plotted in the $M_{DM}-z$ plane for the sake of illustration of the DM search potential with the SKA.\\
\textbf{Dwarf galaxies}, given their extreme proximity, provide an excellent test-bed for DM radio probes, granting access to a parameter space that extends even below the value $\langle \sigma V \rangle = 3 \times 10^{-30}$ cm$^3$ s$^{-1}$. Additionally their large mass-to-light ratios and absence of strong star formation and diffuse non-thermal emission make them very clean sources for radio DM 
searches.\\
\textbf{Galaxies} can be probed to significantly larger redshifts  than the dwarf galaxies due to their larger DM mass, and those located in the redshift range $0.5 \simlt z \simlt 1.0$ provide stronger constraints. However, an optimized DM search should be confined to galaxies with little background radio noise, making low star-formation-rate galaxies good candidates. High-$z$ galaxies come also with the advantage of observing more primitive structures with fewer sources of baryonic radio emission.\\
\textbf{Clusters of galaxies} provide extremely good candidates in cases, such as the Bullet cluster, where the dark and baryonic matter are spatially separated. 
Our recent analysis of the ATCA observation of the Bullet cluster (Colafrancesco and Marchegiani 2014) indicates that deeper radio observations (possible with the SKA) will be able indeed to separate the DM-induced signal from the CR-induced one and hence have the possibly to investigate the nature of DM particles using the  technique here proposed.
More in general, the large predicted radio fluxes due to DM annihilation in clusters indicate that DM-induced radio emission can be observed in radio out to large redshifts $z \approx 2$, again with the advantage of fewer sources of baryonic radio emission.

\subsection{Disentangling magnetic fields and Dark Matter}

Studying the magnetic properties of DM halos are crucial to disentangle the DM particle density from the magnetic field energy density contributing to the expected synchrotron radio emission from DM annihilation.
The SKA is the most promising experiment to determine the magnetic field structure in extragalactic sources 
(see Johnston-Hollitt et al. 2015), and will have the potential of measuring RMs toward a large number of sources  
allowing a detailed description of the strength, structure, and spatial distribution of magnetic fields in dSph galaxies, galaxies (see Beck et al. 2015) and galaxy clusters (see Govoni et al. 2015). 
We stress that these measurements of the magnetic field can be obtained by the SKA simultaneously, for the first time, with the constraints on DM nature from the expected radio emission.

\section{The impact of SKA on the search for the DM nature}

Deep observations of radio emission in DM halos are not yet available, and this limits the capabilities of the current radio experiments to set relevant constraints on DM models.
We have already explored a project (Regis et al. 2014a,b,c)  dedicated to the WIMP search making use of radio interferometers, that could be considered as a pilot experiment for the next generation high-sensitivity and high-resolution radio telescope arrays like the SKA.
For the particle DM search we are interested in, the use of multiple array detectors having synthesized beams of $\sim$ arcmin size has a number of advantages with respect to single-dish observations. First, the large collecting area allows for an increase in the sensitivity over that a single-dish telescope. The best beam choice for the detection of a diffuse emission requires a large synthesized beam (in order to maximize the integrated flux), but still smaller than the source itself to be able to resolve it. 
A good angular resolution is also crucial in order to distinguish between a possible non-thermal astrophysical emission and the DM-induced signal, which clearly becomes very hard if the DM halo is not well-resolved. The possibility of simultaneously detecting small scale sources with the long-baselines of the array allows one to overcome the confusion limit. In the case of arcmin beams, the confusion level can be easily reached with observations lasting for few tens of minutes, even by current
telescopes. A source subtraction is thus a mandatory and crucial step of the analysis. Finally, single dish telescopes
face the additional complication related to Galactic foreground contaminations, which are instead subdominant for
the angular scales typically probed by telescope arrays at GHz frequency. 

The limits derived from ATCA observations of 6 dSphs (Regis et al. 2014c) on the WIMP annihilation/decay rate as a function of the mass for different final states of annihilation/decay are already comparable to the best limits obtained with $\gamma$-ray observations and are much more constraining than what obtained in the X-ray band or with previous radio observations (Spekkens et al. 2013, Natarayan et al. 2013).
In this context, the SKA will have the possibility to explore DM models with cross-section values well below the DM relic abundance one (see Fig.6 in Regis et al. 2014c).
The SKA1-MID Band 1 (350 -1050 MHz) will probably be the most promising frequency range for the majority of WIMP models. The full SKA-2 phase will bring another factor $\sim 10 \times$ increase in sensitivity and an extended frequency range up to at least 25 GHz. 
Typical values of the SKA sensitivity ($A_{eff} /T_{sys} = 2 \times 10^4$ m$^2$/K) and bandwidth ($300$ MHz at
GHz frequency) provide rms flux values of $\approx 30$ nJy for 10 hours of integration time. This is about a $10^3$ factor
of gain in sensitivity with respect to the most recent ATCA observations (Regis et al. 2014a). A further improvement by a factor of 2-3 can be confidently foreseen due to the larger number of accessible dSph satellites from the southern hemisphere.
The SKA will also have the unique advantage to be able to determine the dSph magnetic field (via FR
measurements and possibly also polarization), provided its strength is around the $\mu$G level (as expected
from  star formation rate arguments, Regis et al. 2014c). This will make the predictions for the expected DM signal much
more robust and obtainable with a single experimental configuration.
The prospects of detection/constraints of the WIMP particle properties with the SKA will therefore progressively close in on the full parameter space, even in a pessimistic sensitivity case, and up to $\sim$ TeV WIMP masses, irrespective of astrophysical assumptions.

The SKA will also allow to investigate the possibility that point-sources detected in the proximity of the dSph optical center might be associated to the emission from a DM cuspy profile. This possibility is likely only in the "loss at injection" scenario, while spatial diffusion should in any case flatten the e$^{\pm}$ distribution, making the source extended rather than point-like. 
The investigation of these sources with the SKA will deserve particular attention, since we have already found that the WIMP scenario can fit the point-like emission with annihilation rates consistent with existing bounds (Regis et al. 2014c).

\begin{figure}[ht!]
\centering
\includegraphics[scale=0.38]{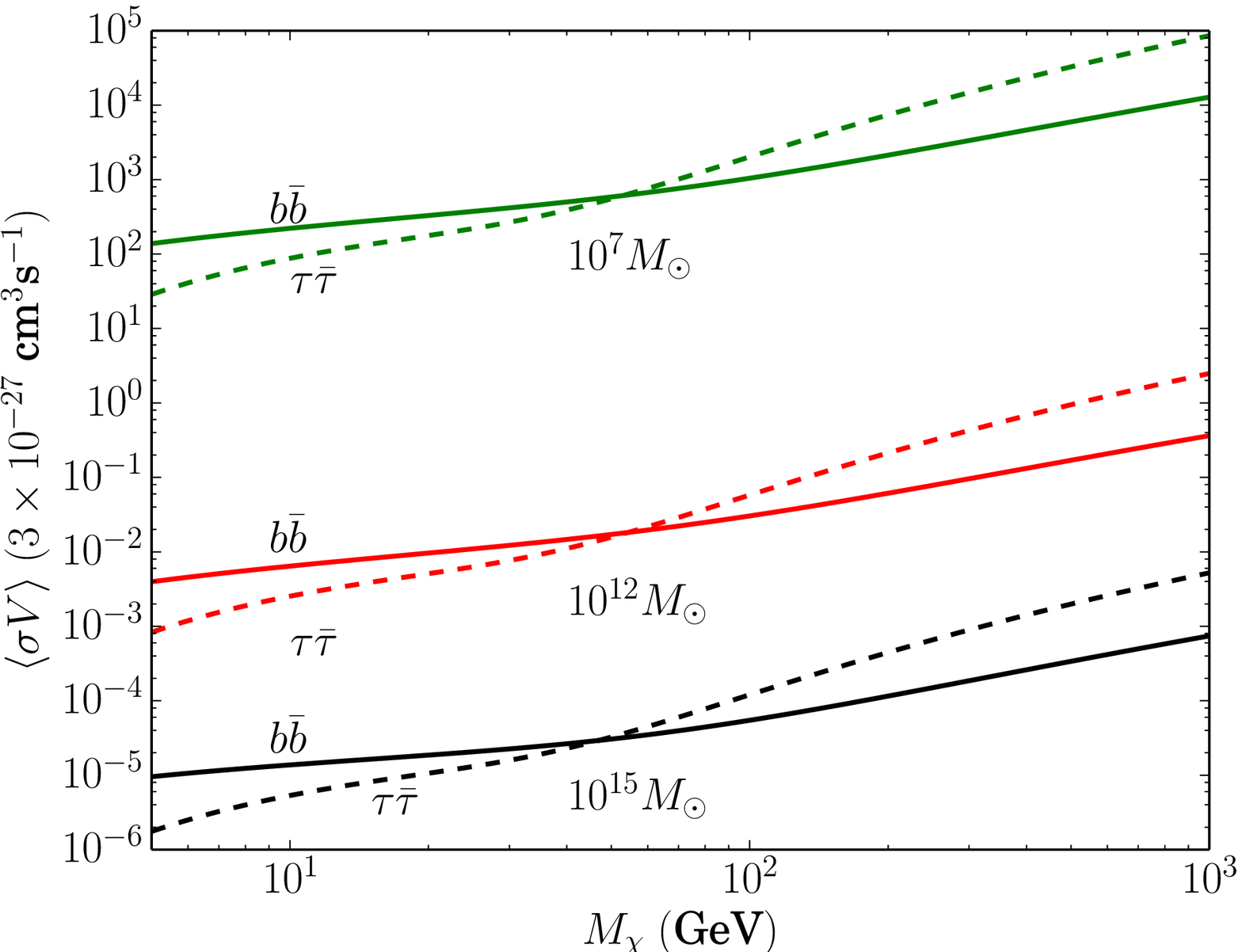}
\includegraphics[scale=0.38]{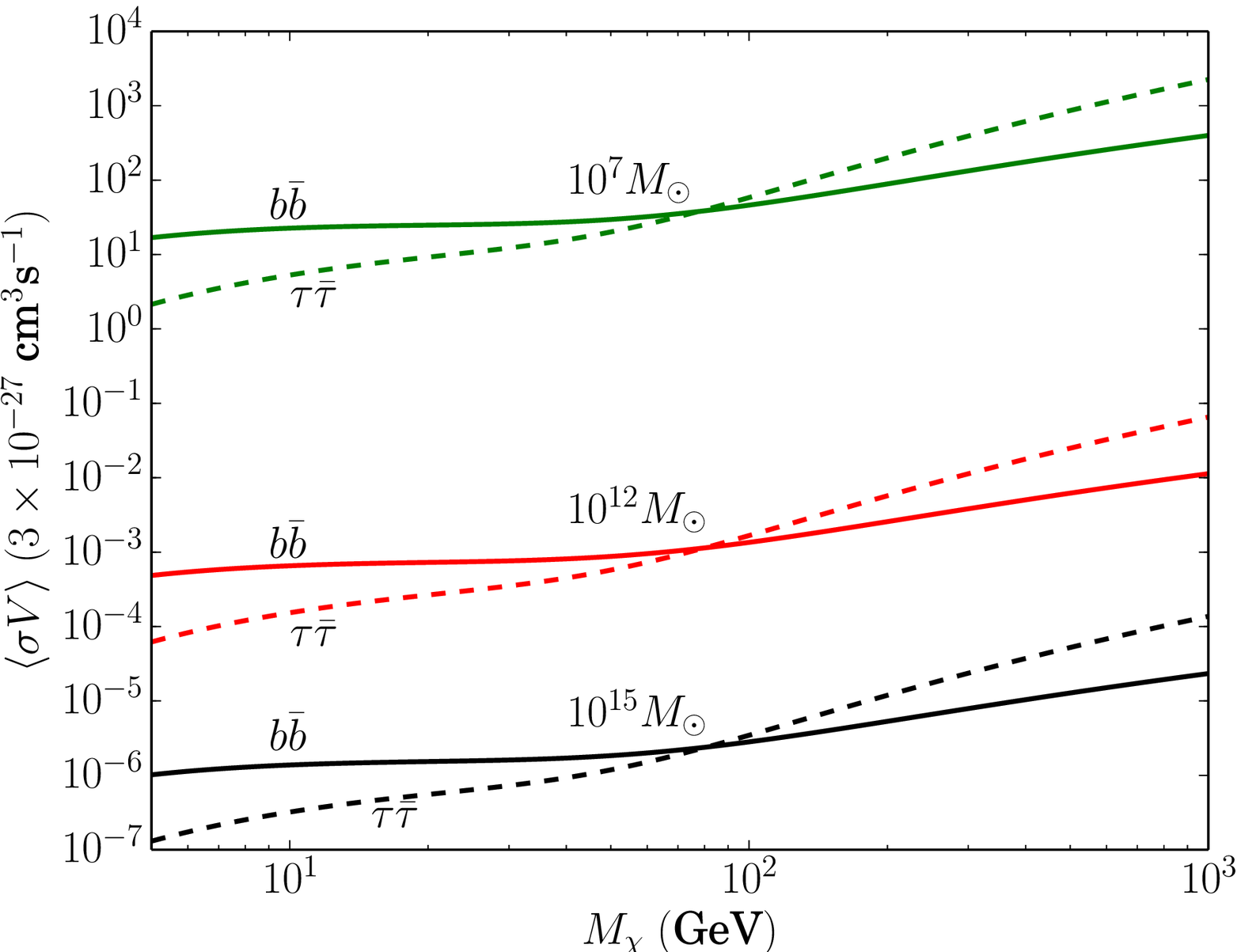}
\caption{The $\langle \sigma V\rangle$ upper limits from 30 hour of SKA integration time for $z = 0.01$ at 300 MHz (top) and 1 GHz (bottom) as the neutralino mass $M_{\chi}$ is varied with annihilation channel $b\overline{b}$ in solid lines and $\tau\bar{\tau}$ in dashed lines. A value $\langle B \rangle =5$ $\mu$G was adopted. Black lines correspond to halos with mass $10^{15}$ M$_{\odot}$, red lines to $10^{12}$ M$_{\odot}$ and green lines to $10^{7}$ M$_{\odot}$ (from Colafrancesco et al. 2014).}
\label{fig:sigv_z0}
\end{figure}


The SKA1-MID Band 1 (350-1050  MHz) to Band 4 (2.8-5.18 GHz) are important to probe the DM-induced synchrotron spectral curvature at low-$\nu$ (sensitive to DM composition) and at high-$\nu$ (sensitive to DM particle mass), and the implementation of Band 5 (4.6-13.8 GHz) will bring further potential to assess the DM-induced radio spectrum. 
As we can see from Fig.\ref{fig:evol_bb60}, the best frequency range to detect these radio emissions is around 1 GHz.
So, the upper frequency regions of SKA1-MID Band 1 provides the strongest spectral candidate for probing the cross-section parameter space due to an optimal combination of the SKA sensitivity within this band and the relatively strong fluxes at these frequencies.
This frequency band will also allow for an optimal description of the magnetic field (see Johnston-Hollitt et al. 2014).



There are two main caveats in the forecasts for DM detection in the radio frequency band. 
The first stems from the fact that, for an extended radio emission,the confusion issue becomes stronger and stronger as one tries to probe fainter and fainter fluxes. Thus, the source subtraction procedure becomes crucial and this can affect the estimated sensitivities. The impact of this effect on the actual sensitivity is hardly predictable at the present time, especially for the SKA, since it will depend on the properties of the detected sources, the efficiency of deconvolution algorithms, and the accuracy of the telescope beam shape.\\
The second caveat is that by bringing down the observational threshold, one can possibly start to probe the very
low levels of possible non-thermal emission associated to the tiny rate of star formation in dSph, or in galaxies and galaxy clusters. 
The DM contribution should be then disentangled from such astrophysical background. 
The superior angular resolution of the SKA will allow for the precise mapping of emissions, putatively either DM or baryonically induced, and will enable their correlation with the stellar or DM profiles (obtained  via optical and/or kinematic measurements).



Early DM science can be done with a small sample of local dSphs, a small sample of nearby galaxies with good DM density profile reconstruction, and the Bullet cluster, observing with a somewhat larger beam ($\approx 7^{{\prime}{\prime}} -10^{{\prime}{\prime}}$).
These objects have been already studied in radio with similar objectives (e.g., DM limits) and therefore provide the best science cases to prove the capabilities of the SKA1 for the study of the nature of DM with radio observations.
The implementation of the SKA1-MID Band 5 (4.6-13.8 GHz) will increase the ability to detect the expected high-frequency spectral curvature of DM-induced radio emissions, and the ability to place constraints on the DM cross-section and to differentiate between different annihilation channel spectra. 
The 10$\times$ increased sensitivity of SKA2-MID compared to SKA1-MID will allow us to increase the angular resolution by a factor $\approx 20$ or to increase the sensitivity of SKA2-MID to DM-induced radio signals by a factor $\approx10$.
The possibility to extend the frequency coverage of the SKA in its Phase-2 realization up to $\sim 25$ GHz, will allow to detect the expected high-$\nu$ spectral cut-off of DM-induced radio emissions, and then set accurate constraints to the DM particle mass.

\section{Conclusion}

The SKA has the potential to unveil the elusive nature of Dark Matter. Its ability to resolve the intrinsic degeneracy (between magnetic field properties and particle distribution) of the synchrotron emission expected from secondary particles produced in DM annihilation (decay) will allow such a discovery to be unbiased and limited only by the sensitivity to the DM particle mass and annihilation cross-section (decay rate). The unprecedented sensitivity of the SKA to the DM fundamental properties will bring this instrument in a leading position for unveiling the nature of the dark sector of the universe.\\
%
The information provided by the SKA can be complemented with analogous studies in other spectral bands, which will be able to prove the ICS signal of DM-produced secondary electrons  (spanning from $\mu$waves to hard X-rays and $\gamma$-rays) and the distinctive presence of the $\pi^0 \to \gamma \gamma$ emission bump in the $\gamma$-rays (see Fig. \ref{fig:multiflux}). 
The next decade will offer excellent multi-frequency opportunities in this respect with the advent of Millimetron, the largest space-borne single-dish mm. astronomy satellite operating in the $10^2-10^3$ GHz range (optimal to prove the DM-induced SZ effect), the Astro-H mission operating in the hard X-rays frequency range (with the highest expected sensitivity  to probe the high-energy tail of the DM-induced ICS emission), and the CTA with unprecedent sensitivity in the energy range between a few tens GeV to hundreds TeV.

\thanks{S.C., P.M. and G.B. acknowledge support by the South African Research
Chairs Initiative of the Department of Science and Technology and National
Research Foundation and by the Square Kilometre Array (SKA).}


\begin{thebibliography}{99}

\bibitem{Ackerman2014} 
Ackermann, M., et al., 2014, Phys. Rev. D, 89, 042001


\bibitem{Beck2014}
Beck, R., Bomans, D., Colafrancesco, S., et~al.\ 2015, in
\emph{Advancing Astrophysics with the Square Kilometre Array}, PoS(AASKA14)94



\bibitem{Borrielloetal2010}
Borriello, E. et al. 2010, ApJ, 709, L32


\bibitem{burns1995-a2255}
Burns, J.O. {\it et al.}., 1995, ApJ, 446,583

\bibitem{Colafrancesco2004}
Colafrancesco, S., 2004, A\&A, 422, L23

\bibitem{Colafrancesco2010} 
Colafrancesco, S. 2010, invited lecture at the 4th Gamow International Conference on Astrophysics \& Cosmology After Gamow (9th Gamow Summer School), AIPC, 1206, 5C


\bibitem{ColafrancescoMele2001} 
Colafrancesco, S. \& Mele, B. 2001, ApJ, 562, 24

\bibitem{CPU2006} 
Colafrancesco, S., Profumo, S. and Ullio, P. 2006, A\&A, 455, 21

\bibitem{CPU2007} 
Colafrancesco, S., Profumo, S. and Ullio, P. 2007, PhRvD, 75, 3513

\bibitem{Colafrancescoetal2011} 
Colafrancesco, S. et al. 2011, A\&A, 527, 80

\bibitem{ColafrancescoMarchegiani2014}
Colafrancesco, S., and Marchegiani, P, 2014, A\&A in press

\bibitem{Colafrancescoetal2014}
Colafrancesco, S., Marchegiani, P. and Beck, G., 2014, [2014arXiv1409.4691C]

\bibitem{Daylanetal2014} 
Daylan, T. et al., 2014, [arXiv:1402.6703]

\bibitem{Deissetal1997}
Deiss, B.M., Reich, W., Lesch, H. \& Wielebinski, R. 1997, A\&A, 321, 55

\bibitem{ska2012} 
Dewdney, P., Turner, W., Millenaar, R., McCool, R., Lazio, J. \& Cornwell, T., 2012, SKA baseline design document, {\slshape http://www.skatelescope.org/wp-content/uploads/2012/07/SKA-TEL-SKO-DD-001-1\_BaselineDesign1.pdf}

\bibitem{Doro2014}
Doro, M., 2014, Nuclear Instruments and Methods in Physics Research A,742, 99

\bibitem{Doro2013} 
Doro, M., et al., 2013, Astroparticle Physics, 43, 189




\bibitem{Govonietal2006}
Govoni, F., Murgia, M., Feretti, L., et al., 2006, A\&A, 460, 425

\bibitem{Govoni2014}
Govoni, F. et~al.\ 2015, in
\emph{Advancing Astrophysics with the Square Kilometre Array}, PoS(AASKA14)105


\bibitem{JH2014}
Johnston-Hollitt, M. et~al.\ 2015, in
\emph{Advancing Astrophysics with the Square Kilometre Array}, PoS(AASKA14)092

\bibitem{Jungmanetal1996} 
Jungman, G., Kamionkowki, M. \& Griest, K. 1996, Phys.rep, 267, 195







\bibitem{Moranetal2014}
Moran, E.C. et al. 2014, [arXiv:1408.4451]

\bibitem{GBT_2}
Natarajan, A. et al., 2013, Phys. Rev. D 88 083535 [arXiv:1308.4979 [astro-ph.CO]].



\bibitem{PlanckSZ2011}
Planck Collaboration, 2011, A\&A, 536, A7

\bibitem{Regis2014_1}
Regis, M., Richter, L., Colafrancesco, S., Massardi, M., de Blok, W. J. G., Profumo, S., Orford, N. 2014a, [arXiv:1407.5479]

\bibitem{Regis2014_2}
Regis, M., Richter, L., Colafrancesco, S., Profumo, S., de Blok, W. J. G., Massardi, M. 2014b, [arXiv:1407.5482]

\bibitem{Regis2014_3}
Regis, M., Colafrancesco, S., Profumo, S., de Blok, W. J. G., Massardi, M., Richter, L. 2014c, JCAP, 10, 016R [arXiv:1407.4948]

\bibitem{GBT_1}
Spekkens, K., Mason, B.S., Aguirre, J.E. and Nhan, B., 2013, ApJ, 773, 61 [arXiv:1301.5306 [astro-ph.CO]].


\bibitem{Weniger2012} 
Weniger, C., 2012, JCAP, 1208, 007


\end{thebibliography}
\end{document}